\pdfoutput=1
\documentclass[compress]{elsarticle}
\makeatletter
\def\ps@pprintTitle{%
 \let\@oddhead\@empty
 \let\@evenhead\@empty
 \def\@oddfoot{\centerline{\thepage}}%
 \let\@evenfoot\@oddfoot}
\makeatother

\usepackage{todonotes}
\usepackage{float}
\biboptions{sort&compress}
\usepackage{mathtools}
\usepackage{graphicx}
\usepackage{flushend}
\usepackage{float}
\usepackage{grffile}
\usepackage{amsmath}
\usepackage[inline]{enumitem}
\usepackage{enumitem}

\usepackage{amsthm}
\usepackage{hyperref}
\newcommand{\vbyte}{VByte}
\newcommand{\maskedvbyte}{\textsc{Masked \vbyte{}}}
\newcommand{\varintgb}{\textsc{varint-GB}}
\newcommand{\streamvbyte}{\textsc{Stream \vbyte{}}}
\newcommand{\varintgiu}{\textsc{varint-G8IU}}
\usepackage[utf8x]{inputenc}
\usepackage{booktabs}
\usepackage{listings}
\usepackage{algorithm}
\usepackage[noend]{algorithmic}
\usepackage{subfloat,subfig}
\usepackage{tikz}
\usepackage{color}
\usetikzlibrary{chains,positioning,arrows}

\usepackage{times}

\usepackage{siunitx}

\graphicspath{{./results/}}
\usepackage{url}

\newcommand{\ourtitle}{\streamvbyte{}: Faster Byte-Oriented Integer Compression}
\long\def\ignore#1{}

\journal{Information Processing Letters}
 \author[UQAM]{Daniel Lemire\corref{cor1}} \ead{lemire@gmail.com}
 \author[verse]{Nathan Kurz} \ead{natekurz@gmail.com}
 \author[Upscaledb]{Christoph Rupp} \ead{chris@crupp.de}

 \address[UQAM]{\scriptsize LICEF, Universit\'e du Qu\'ebec,
 5800 Saint-Denis, Montreal, QC, H2S 3L5 Canada
}
\address[verse]{\scriptsize Orinda, California USA}
\address[Upscaledb]{\scriptsize Upscaledb, Munich, Germany}

 \cortext[cor1]{Corresponding author. Tel.: 00+1+514 843-2015
   ext. 2835; fax: 00+1+800 665-4333.}

\begin{document}
\title{\ourtitle{}}

\begin{frontmatter}

\begin{abstract}
Arrays of integers are often compressed
in search engines. Though there are many ways to compress
integers, we are interested in the popular byte-oriented integer compression
techniques (e.g., \vbyte{} or Google's \varintgb{}). Although not known for their speed, they are appealing due to their simplicity and engineering convenience.
Amazon's \varintgiu{} is one of the fastest
byte-oriented compression technique published so far.
It makes judicious use of the powerful single-instruction-multiple-data (SIMD) instructions
available in commodity processors. To surpass \varintgiu{},
we present \streamvbyte{}, a novel byte-oriented compression
technique that separates the control stream from the encoded data. Like \varintgiu{}, \streamvbyte{} is well suited for SIMD instructions.
 We show that \streamvbyte{} decoding can be up to twice as fast as  \varintgiu{} decoding over real data sets. In this sense,
\streamvbyte{} establishes new speed records for byte-oriented integer compression, at times exceeding the speed of the \texttt{memcpy} function. On a \SI{3.4}{GHz}  Haswell processor, it decodes more than 4~billion  differentially-coded integers per second from RAM to L1 cache.

\end{abstract}

\begin{keyword}
Data compression \sep Indexing \sep Vectorization \sep SIMD Instructions
\end{keyword}

\end{frontmatter}

\section{Introduction}

We frequently represent sets of document or row identifiers by arrays of integers.
Compressing these arrays can keep the data closer to
the processor and reduce bandwidth usage, and fit more data in memory or on a disk.
Though data can always be compressed using generic algorithms such as
Lempel-Ziv coding (LZ77), specialized compression algorithms
for integers can be orders of magnitude faster.
We consider codecs to compress 32-bit unsigned integers (in $[0, 2^{32})$).
 We are especially interested in compressing arrays where most integers
are small. In addition to such arrays that arise naturally, sorted arrays of non-small integers can often be treated as arrays of small integers by considering the  successive differences (``deltas''). So 
instead of compressing the integers $x_1, x_2, \ldots$ directly, we can compress the gaps between them
$x_1, x_2-x_1, x_3-x_2, \ldots$
There are many integer-compression algorithms applicable to
arrays of small integers.
One option are byte-oriented techniques~\cite{Stepanov:2011:SDP:2063576.2063627}. In these formats, the main data
corresponding to an integer is stored in
consecutive whole bytes, and all bits within a given byte correspond to only one such integer.
Though byte-oriented formats do not offer the best compression ratios or the best speeds, they are in widespread use within databases, search engines and data protocols in part because of their simplicity. 

Historically, the most widely known byte-oriented compression algorithm is \vbyte{}.
It writes a non-negative integer starting from the least significant bits, using  seven bits in each byte, with the most significant bit set to zero when the following byte continues the current integer.   
Thus integers in  $[0,2^7)$ are coded using a single byte, integers in $[2^7, 2^{14})$ in two bytes and so on.
For example, the integer 32 is written using a single byte (\textbf{0}0100000)
and the integer 128 is written using two~bytes (\textbf{1}0000000 and \textbf{0}0000001).

To decode \vbyte{} data, it suffices to iterate
over the compressed bytes while checking the value of the most significant bit.
Whenever the number of bytes required per integer is easily predictable---such as when most integers fit in $[0,2^7)$---the absence of branch prediction errors allows high decoding speeds on modern-day superscalar processors capable of speculative execution.

 Unfortunately,
not all data is easily predictable.
In some of these case, the decoding speed of \vbyte{} compressed data becomes a bottleneck.
For this reason, Google developed \varintgb{}~\cite{DeanOfficialplusslides:2009:CBL:1498759.1498761}. In \varintgb{}, numbers are compressed and decompressed in blocks of four.
Instead of having to check one
bit per compressed byte, only one control byte for every four integers needs to be processed.
For unpredictable patterns, this block-based design can reduce the number of branch mispredictions by a factor of four.

Modern processors  (e.g., ARM, POWER, Intel, AMD) have instructions 
that perform the same operation on multiple scalar values  (e.g., the addition of two sets of four 32-bit integers).
These instructions are said to be  \emph{single-instruction-multiple-data} (SIMD) while the algorithms designed to take advantage of SIMD instructions are said to be \emph{vectorized}.

 Stepanov et al.~\cite{Stepanov:2011:SDP:2063576.2063627} considered the problem of vectorizing byte-oriented compression.
  They reported that there was little benefit to vectorizing VByte  decoding.  As an alternative to \vbyte{} and \varintgb{}, Stepanov et al.\ proposed a new patented
 byte-oriented format (\varintgiu{}) designed with SIMD instructions in mind. It proved more than three times faster than \vbyte{} on realistic sets of document identifiers.

Plaisance et al.~\cite{plaisance2015} revisted  the \vbyte{} decoding problem: unlike Stepanov et al., their  \maskedvbyte{} decoder is twice as fast as  a scalar \vbyte{}  decoder.
Though it  should not be expected to be as fast as a decoder working on a format designed for SIMD instructions (e.g., \varintgiu{}), it can
help in systems where the data format is fixed.

There remained an open question: could \varintgiu{} be surpassed?
Could byte-oriented decoding be even faster? We answer this question
by the positive. In fact, our proposal (\streamvbyte{}) can be twice as fast as \varintgiu{} on realistic data.

\section{SIMD Instructions}
\label{sec:simdinst}

Since the introduction of the Pentium~4 in 2001, x64 processors have had  vector instructions operating on 16-byte SIMD registers (called XMM registers). These
registers  can store four 32-bit integers.

A \streamvbyte{} decoder (\S~\ref{sec:streamvbyte}) can be written with  just two x64 SIMD assembly instructions:
\begin{itemize}[noitemsep,nolistsep,leftmargin=10pt]
\item The \texttt{movdqu} instruction can load 16~bytes from memory into a SIMD register, or write such a register to memory. On recent Intel processors (e.g., Haswell) these operations have multicycle latency (5--6~cycles from L1 cache), but they also have high   throughput.
In ideal cases, two XMM loads and an XMM write can all be issued each CPU cycle.
\item The shuffle (\texttt{pshufb}) instruction  can selectively copy the
byte values of one SIMD register  $v$ to another according to a mask $m$.
If $v_0, v_1, \ldots, v_{15}$ are the values of the 16~individual bytes in $v$,  and $m_0, m_1, \ldots, m_{15}$ are the bytes within $m$ ($m_i\in\{-1,0,1,2,\ldots, 15\}$), then \texttt{pshufb} outputs $(v_{m_0},v_{m_1},\ldots, v_{m_{15}})$ where $v_{-1}\equiv 0$. Once its operands are in registers, the  \texttt{pshufb} instruction is fast: it has a latency of one cycle and a reciprocal throughput of one instruction per cycle.
\end{itemize}
Both of these SIMD instructions are available in all processors supporting the SSSE3 instruction set, i.e., almost all x64 processors produced since 2010.

\paragraph{Differential Coding} When we have compressed ``deltas'', or successive differences $\delta_1, \delta_2, \ldots = x_1, x_2-x_1, x_3-x_2, \ldots$
 instead of the original integers, we need to compute
 a prefix sum to recover the original integers ($\delta_1, \delta_1+\delta_2, \delta_1 + \delta_2 + \delta_3, \ldots$).
The computation of the prefix sum can be accelerated by vector shifts and additions.
We can compute the prefix sum of a vector of size $2^L$ using $L$~shifts and $L$~additions~\cite{SPE:SPE2326}. For example, consider the vector of 
delta values $(3,4,12,1)$. We add to this vector a version of itself shifted by one integer ($(0,3,4,12)$) to get $(3,7,16,13)$. Finally, we add to this last vector a version of itself shifted by two integers ($0,0,3,7$) to get the prefix sum $(3,7,19,20)$ of the original vector.

\section{Byte-Oriented Integer Codecs}

The  \vbyte{} format uses the most significant bit of each
byte as a control bit.
Thus an integer is compressed to an array of $L+1$~bytes
 $b_0, b_1, \ldots, b_L $ such that
$b_0, b_1, \ldots, b_{L-1}$ have their most significant bits set
to 1 whereas the most significant bit of $b_L$ is zero.
That is, if we view bytes as unsigned integers in $[0,2^8)$, we have
that $b_0, b_1, \ldots, b_L \in [2^7, 2^8)$ and $b_L\in [0,2^7)$.
 We can decode the compressed integer as  $\sum_{i=0}^L (b_i \bmod 2^7) \times 2^{7 i}$.
 Integers in  $[2^{7(L-1)},2^{7L})$ are coded using  $L$~bytes for $L=1,2,\ldots$
The software to decode the \vbyte{} format is easy to write in standard C, but performance on inputs with mixed byte lengths $L$ can suffer due to poor branch prediction.

For greater speed,
Plaisance et al.~\cite{plaisance2015} proposed the \maskedvbyte{} decoder that uses SIMD instructions. It works directly on
the standard \vbyte{} format.
 The \maskedvbyte{} decoder gathers the most significant bits  of an array of consecutive bytes using the \texttt{pmovmskb} x64/SSE2 instruction. Using look-up tables and a shuffle instruction (\texttt{pshufb}),  \maskedvbyte{} permutes the bytes  to arrive at the decoded integers. We refer the interested reader to
Plaisance et al.~\cite{plaisance2015} for a detailed description.


\begin{table}[b]
\caption{Overview of the byte-oriented integer compression techniques\label{sec:overview}}
\small
\begin{tabular}{cp{6.9cm}c}
\toprule
name of decoder & data format   & SIMD \\
        \midrule
 \vbyte{}       & 7~data~bits per byte, 1 bit as continuation flag      &  no \\
\maskedvbyte{}~\cite{plaisance2015}  & identical to \vbyte{} &  yes \\
\varintgb{}~\cite{DeanOfficialplusslides:2009:CBL:1498759.1498761}  &
fixed number of integers (4) compressed to a variable number of bytes (4--16), prefixed by a control byte  &  no \\
\varintgiu{}~\cite{Stepanov:2011:SDP:2063576.2063627}  &
fixed number of compressed bytes (8) for a variable number of integers (2--8), prefixed by a control byte      &  yes \\
\streamvbyte{} (novel) &  control bytes and data bytes in separate streams &  yes \\\bottomrule
\end{tabular}
\end{table}

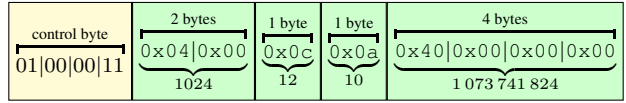
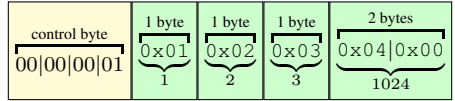
\begin{figure}
\centering
\subfloat[Integer values $1024,12,10,\num{1073741824}$ compressed using 9~bytes]{
\centering\begin{tikzpicture}[node distance=0cm,start chain=1 going right]  \footnotesize
 \tikzstyle{mytape}=[draw,minimum height=1.3cm]
    \node  [on chain=1,mytape,fill=yellow!20] {$\overbracket{01|00|00|11}^{\text{control byte}}$};
    \node  [on chain=1,mytape,fill=green!20] {$\underbrace{\overbracket{\texttt{0x04}|\texttt{0x00}}^{\text{2 bytes}}}_{1024}$};
    \node  [on chain=1,mytape,fill=green!20] {$\underbrace{\overbracket{\texttt{0x0c}}^{\text{1 byte}}}_{12}$};
    \node  [on chain=1,mytape,fill=green!20] {$\underbrace{\overbracket{\texttt{0x0a}}^{\text{1 byte}}}_{10}$};
    \node  [on chain=1,mytape,fill=green!20] {$\underbrace{\overbracket{\texttt{0x40}|\texttt{0x00}|\texttt{0x00}|\texttt{0x00}}^{\text{4 bytes}}}_{\num{1073741824}}$};
\end{tikzpicture}
}

\subfloat[Integer values $1,2,3,1024$ using 6~bytes]{
\centering\begin{tikzpicture}[node distance=0cm,start chain=1 going right] \footnotesize
  \tikzstyle{mytape}=[draw,minimum height=1.3cm]
    \node  [on chain=1,mytape,fill=yellow!20] {$\overbracket{00|00|00|01}^{\text{control byte}}$};
    \node  [on chain=1,mytape,fill=green!20] {$\underbrace{\overbracket{\texttt{0x01}}^{\text{1 byte}}}_{1}$};
    \node  [on chain=1,mytape,fill=green!20] {$\underbrace{\overbracket{\texttt{0x02}}^{\text{1 byte}}}_{2}$};
    \node  [on chain=1,mytape,fill=green!20] {$\underbrace{\overbracket{\texttt{0x03}}^{\text{1 byte}}}_{3}$};
    \node  [on chain=1,mytape,fill=green!20] {$\underbrace{\overbracket{\texttt{0x04}|\texttt{0x00}}^{\text{2 bytes}}}_{\num{1024}}$};
\end{tikzpicture}
}
\caption{\label{fig:groupvarint}Examples of blocks of four integers  compressed with \varintgb{}.}
\end{figure}

 Whereas
the  \vbyte{} format interleaves actual integer data (using the least significant 7~bits of each byte) with control data (using the most significant bit of each byte), \varintgb{} stores the control
data using a distinct control byte corresponding to a block of four~integers~\cite{DeanOfficialplusslides:2009:CBL:1498759.1498761}.
See Table~\ref{sec:overview} and Fig.~\ref{fig:groupvarint}.
To store an integer $x$, we use  $\lceil \log_{2^8} (x + 1) \rceil $~continuous bytes, since
$x \in [0,2^{\lceil \log_{2^8} (x + 1) \rceil})$ always holds.
 Given four integers, $x_1, x_2, x_3, x_4$, the control
byte stores the lengths $\lceil \log_{2^8} (x_i + 1) \rceil$ for each of the four integers $i=1,2,3,4$. Indeed, because there are only four possible byte~lengths ($\{1,2,3,4\}$), each of the four individual byte lengths can be stored using just two bits.
Thus a block in \varintgb{} used to store integers $x_1, x_2, x_3, x_4$ contains a control byte followed
by  $\sum_{i=1}^4 \lceil \log_{2^8} (x_i + 1) \rceil$~bytes.
When the number of integers is not divisible by four, an incomplete block might be used. For this reason, it might be necessary to record how many integers are compressed  (e.g., as a format header).

To decode \varintgb{} data, the control byte is first parsed, from which we extract the four lengths.
The integers $x_1, x_2, x_3, x_4$ are then extracted (without any branching) and we advance to the next control byte. To achieve higher speed for highly compressible data, we find it useful to include an optimized code path for when all integers are small ($x_1, x_2, x_3, x_4 \in [0,2^8)$).

\begin{figure}
\centering
\subfloat[Integer values $1024,12,10,\num{1073741824}$ compressed using 9~bytes]{
\centering\begin{tikzpicture}[node distance=0cm,start chain=1 going right]  \footnotesize
 \tikzstyle{mytape}=[draw,minimum height=1.3cm]
    \node  [on chain=1,mytape,fill=yellow!20] {$\overbracket{10|0|0|1110}^{\text{control byte}}$};
    \node  [on chain=1,mytape,fill=green!20] {$\underbrace{\overbracket{\texttt{0x04}|\texttt{0x00}}^{\text{2 bytes}}}_{1024}$};
    \node  [on chain=1,mytape,fill=green!20] {$\underbrace{\overbracket{\texttt{0x0c}}^{\text{1 byte}}}_{12}$};
    \node  [on chain=1,mytape,fill=green!20] {$\underbrace{\overbracket{\texttt{0x0a}}^{\text{1 byte}}}_{10}$};
    \node  [on chain=1,mytape,fill=green!20] {$\underbrace{\overbracket{\texttt{0x40}|\texttt{0x00}|\texttt{0x00}|\texttt{0x00}}^{\text{4 bytes}}}_{\num{1073741824}}$};
\end{tikzpicture}
}

\subfloat[Integer values $1,2,3,1024, 1024$ compressed using 9~bytes \label{fig:waste}]{
\centering\begin{tikzpicture}[node distance=0cm,start chain=1 going right] \footnotesize
  \tikzstyle{mytape}=[draw,minimum height=1.3cm]
    \node  [on chain=1,mytape,fill=yellow!20] {$\overbracket{0|0|0|10|10|\underbrace{\color{red}1}_{\text{\color{red}waste}}}^{\text{control byte}}$};
    \node  [on chain=1,mytape,fill=green!20] {$\underbrace{\overbracket{\texttt{0x01}}^{\text{1 byte}}}_{1}$};
    \node  [on chain=1,mytape,fill=green!20] {$\underbrace{\overbracket{\texttt{0x02}}^{\text{1 byte}}}_{2}$};
    \node  [on chain=1,mytape,fill=green!20] {$\underbrace{\overbracket{\texttt{0x03}}^{\text{1 byte}}}_{3}$};
    \node  [on chain=1,mytape,fill=green!20] {$\underbrace{\overbracket{\texttt{0x04}|\texttt{0x01}}^{\text{2 bytes}}}_{\num{1024}}$};
    \node  [on chain=1,mytape,fill=green!20] {$\underbrace{\overbracket{\texttt{0x04}|\texttt{0x01}}^{\text{2 bytes}}}_{\num{1024}}$};
    \node  [on chain=1,mytape,fill=green!20] {$\underbrace{\overbracket{\color{red}\texttt{0x00}}^{\text{1 byte}}}_{\text{\color{red}waste}}$};

\end{tikzpicture}
}
\caption{\label{fig:varintgiu}Examples of blocks of 9~compressed bytes representing different numbers of integers in the \varintgiu{} format}
\end{figure}
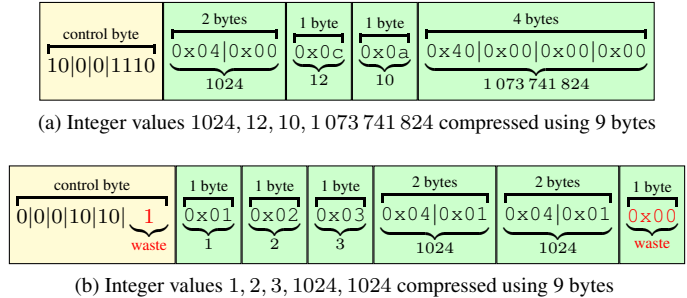

Like \varintgb{}, \varintgiu{} also uses a control byte~\cite{Stepanov:2011:SDP:2063576.2063627},
but it does not describe a fixed number (four) of integers.  Rather, \varintgiu{} divides the compressed output into blocks of eight data bytes for each control byte. Each block can contain between two and eight compressed integers. See Fig.~\ref{fig:varintgiu}.
The control byte describes the next eight~data bytes. Thus a compressed block in \varintgiu{} is always made of exactly nine~compressed bytes.
Each bit in the control byte corresponds to one of the eight data bytes. A bit value of 0~indicates that the data byte completes a compressed integer, whereas a bit value of 1~indicates that the data byte is part of a compressed byte or is ``wasted''. Indeed, as shown in Fig.~\ref{fig:waste}, if the control byte ends with trailing ones, then they correspond to bytes that are wasted (they do not correspond to a compressed integer). A case with waste like the example of Fig.~\ref{fig:waste} might arise if the next integer to be coded is larger or equal to $2^{8}$ so that it cannot fit in the remaining byte.

 \varintgiu{} can sometimes have worse compression than
\vbyte{} due to the wasted bytes. However,
\varintgiu{} can compress a stream of integers in $[2^7,2^8)$ using
only nine~bits per integer on average, against $16$~bytes for \vbyte{}, so \varintgiu{} can also, in principle, compress better. 
\varintgb{} has a slight size disadvantage compared to \varintgiu{} when compressing streams of integers in $[0,2^8)$ as it uses ten~bits per integer against only nine~bits per integer for \varintgiu{}. However,
this is offset by  \varintgb{}'s better compression ratio for larger integers.

\section{\label{sec:streamvbyte}The \streamvbyte{} format}

 Stepanov et al.~\cite{Stepanov:2011:SDP:2063576.2063627} attempted to
 accelerate \varintgb{} by using SIMD instructions. The results were far inferior to \varintgiu{}---despite the apparent
 similarity between the formats, with each having one control byte followed by some data.

 To understand why it might be difficult to accelerate the decoding of
 data compressed in the \varintgb{} format compared to the \varintgiu{}
 format, consider that we cannot decode faster than we can
 access the control bytes. In \varintgiu{}, the control bytes are
 conveniently always located nine~compressed bytes apart. Thus while a control byte
 is being processed, or even before, our superscalar processor can load
 and start
 processing upcoming control bytes, as their locations are predictable.
 Instructions depending on these control bytes can be reordered by the
 processor for best performance.
However, in the \varintgb{} format, there is a strong data dependency: the
 location of the next control byte depends on the current control byte.
 This increases the risk that the processor remains
 underutilized, delayed by the latency
 between issuing the load for the next control byte and waiting for it to be ready.

But the \varintgiu{} format has its own downside: it decodes a variable
number of integers per control bytes (between two and eight inclusively). We expect that it is faster to store full SIMD registers (e.g., four integers) to memory with each iteration. Moreover, when using differential coding,  it is more convenient and efficient to reconstruct the original integers from their deltas when they come in full SIMD registers.

\begin{figure}
\centering
\centering\begin{tikzpicture}[node distance=0cm,start chain=1 going right,start chain=2 going right] \footnotesize
  \tikzstyle{mytape}=[draw,minimum height=1.3cm]
    \node(BIG1)  [on chain=1,mytape,fill=yellow!20] {$\overbracket{01|00|00|11}^{\text{control byte}}$};
    \node(Z2)  [on chain=1,mytape,fill=yellow!35] {$\overbracket{00|00|00|01}^{\text{control byte}}$};

                \node [right of=Z2,node distance=4cm,fill=blue!10] {\parbox{6cm}{\footnotesize Control bytes are stored continuously in a separate address than from the data bytes that are also stored continuously. This layout minimizes latency while accessing the control bytes.}};

    \node(A1)  [on chain=2,mytape,fill=green!20,below of=BIG1,node distance=2cm] {$\underbrace{\overbracket{\texttt{0x04}|\texttt{0x00}}^{\text{2 bytes}}}_{1024}$};
    \node(A2)  [on chain=2,mytape,fill=green!20] {$\underbrace{\overbracket{\texttt{0x0c}}^{\text{1 byte}}}_{12}$};
    \node(A3)  [on chain=2,mytape,fill=green!20] {$\underbrace{\overbracket{\texttt{0x0a}}^{\text{1 byte}}}_{10}$};
    \node(A4)  [on chain=2,mytape,fill=green!20] {$\underbrace{\overbracket{\texttt{0x40}|\cdots|\texttt{0x00}}^{\text{4 bytes}}}_{\num{1073741824}}$};
    \node(B1)  [on chain=2,mytape,fill=green!35] {$\underbrace{\overbracket{\texttt{0x01}}^{\text{1 byte}}}_{1}$};
    \node(B2)  [on chain=2,mytape,fill=green!35] {$\underbrace{\overbracket{\texttt{0x02}}^{\text{1 byte}}}_{2}$};
    \node(B3)  [on chain=2,mytape,fill=green!35] {$\underbrace{\overbracket{\texttt{0x03}}^{\text{1 byte}}}_{3}$};
    \node(B4)  [on chain=2,mytape,fill=green!35] {$\underbrace{\overbracket{\texttt{0x04}|\texttt{0x00}}^{\text{2 bytes}}}_{\num{1024}}$};

    \draw [fill=red!20]  (BIG1.south) --  (A1.north) --  (A4.north) -- cycle;

    \draw [fill=red!35]  (Z2.south) --  (B1.north) --  (B4.north) -- cycle;
    
\draw [-] (BIG1.south) -- (A1.north);
\draw [-] (BIG1.south) -- (A4.north);

\draw [-] (Z2.south) -- (B1.north);
\draw [-] (Z2.south) -- (B4.north);

\end{tikzpicture}
\caption{\label{fig:streamvbyte}Compressed \streamvbyte{} bytes
for $1024,12,10,\num{1073741824}, 1,2,3,1024$.}
\end{figure}

Thankfully, we can combine the benefits of the \varintgb{} and the
\varintgiu{} formats: (1)~having control bytes at predictable locations
so that the processor can access series of control bytes simultaneously,
without data dependency and (2)~decoding integers in full
SIMD~registers.
We use a format that is identical to that of \varintgb{} with the
small, but important, difference that the control bytes are
stored continuously in a separate stream from the data bytes.
See Fig.~\ref{fig:streamvbyte}.
Since we record how many integers $N$ are compressed, we can
 use the first $\lceil  2N / 8  \rceil $~compressed
bytes to store the control bytes followed by the data bytes.

\lstdefinestyle{customc}{%
  belowcaptionskip=1\baselineskip,
  breaklines=true,
  numberstyle=\big\color{blue},
  xleftmargin=\parindent,
  language=C,
  showstringspaces=false,
  basicstyle=\footnotesize\ttfamily,
  keywordstyle=\bfseries\color{green!40!black},
  commentstyle=\small\itshape\color{purple!40!black},
  identifierstyle=\bfseries\color{black},
  stringstyle=\color{orange},
   morekeywords={uint8_t, __m128i,__m256i,__m128i,UINT64_C},
    tabsize=1,
    numbers=left, numberstyle=\bfseries\color{red!30!black}, stepnumber=1, numbersep=5pt
}
\lstset{escapechar=@,style=customc}

\lstset{escapeinside={(*@}{@*)}}
\begin{figure}[b]\centering
\begin{tabular}{c}
\begin{lstlisting}
// "databytes" is a byte pointer to compressed data
// "control" contains control byte
uint8_t C = lengthTable[control]; // C is between 4 and 16 (*@\label{line:step1}@*)
__m128i Data = _mm_loadu_si128((__m128i *) databytes); (*@\label{line:step2}@*)
__m128i Shuf = _mm_loadu_si128(shuffleTable[control]); (*@\label{line:step3}@*)
Data = _mm_shuffle_epi8(Data, Shuf); (*@\label{line:step4}@*) // final decoded data
datasource += C; (*@\label{line:step5}@*)
\end{lstlisting}
\end{tabular}
\caption{\label{fig:code}Core of the \streamvbyte{} decoding procedure in C with Intel intrinsics}
\end{figure}

Decoding a block requires no more than a handful of lines of code (see Fig.~\ref{fig:code}).
At all times, we maintain a pointer into the stream of control bytes and a pointer into the stream of data bytes. They are initialized to the respective beginning of their streams.
\begin{itemize}[noitemsep,nolistsep,leftmargin=10pt]
\item We start by retrieving the control byte.
 Given the control byte, we load from a 256-integer look-up table the number $C$ of corresponding  compressed bytes ($C\in [4,16]$). See Fig.~\ref{fig:code}, line~\ref{line:step1}. For example, if  the control byte is made of only zeros, then  the sought-after length is four.
\item Simultaneously, we load the next 16~data bytes in a 16-byte SIMD register. See Fig.~\ref{fig:code}, line~\ref{line:step2}. Depending on the control byte, we use only $C$ of these bytes.
\item From the control byte, we load a 16-byte shuffling mask for the \texttt{pshufb} instruction (line~\ref{line:step3}). There are 256~such masks, one for each possible value of the control byte.
\item We apply the \texttt{pshufb} instruction on the data SIMD register using the shuffling mask  (line~\ref{line:step4}). The result contains the uncompressed integers. If differential coding is used, the result contains four deltas which can be decoded (\S~\ref{sec:simdinst}).
\item Both pointers are advanced. The control-byte pointer is advanced by one byte whereas the data-byte pointer is advanced by $C$~bytes (line~\ref{line:step5}).
\end{itemize}
Incomplete blocks (containing fewer than four integers) are decoded using a scalar function similar to that used for \varintgb{}. Likewise, when we detect that fewer than 16~data bytes remain, we use a scalar function. Further, we found it useful to use an optimized code path when
we have four~zero control bytes in sequence.

When the data is highly compressible,  the \streamvbyte{}
format stores long runs of control bytes set to zero. In some applications, it might be beneficial
to compress such runs to improve compression (e.g., using run-length encoding).

Though details are outside our scope, we have implemented fast functions
to append new integers to a compressed \streamvbyte{} array without having to recompress the data.
We  append extra data, and occasionally grow the control bytes stream.

\section{Experiments}

We implemented our compression software in C. We use
the GNU GCC~4.8 compiler with  the \texttt{-O3} flag. To ease reproducibility, our software is  available online.\footnote{\url{https://goo.gl/6Op1t4} and \url{https://github.com/lemire/streamvbyte}}

We run the benchmark program on  a Linux server with an Intel i7-4770 processor (\SI{3.4}{GHz}).
This Haswell processor has \SI{32}{kB} of L1 data cache and \SI{256}{kB} of L2 cache per core with \SI{8}{MB} of shared L3 cache.
The machine has ample memory (\SI{32}{GB} of dual-channel DDR3~1600~RAM).
Turbo Boost and Speed~Step are disabled, so the processor runs consistently at its rated speed of \SI{3.4}{GHz}.
We measure wall-clock  times to decompress data from RAM to L1 cache.  All tests are single-threaded.

Search engines typically rely on posting lists: given a term,
we create a list of all document identifiers corresponding to documents where the term appears. 
Instead of directly storing the identifiers, we
use differential coding to reduce the average number of bits required to store each compressed integer. The document identifiers are sorted in
increasing order ($x_1, x_2, \ldots$ where $x_i>x_{i-1}$ for $i=2,3,\ldots$),
and we compress their successive differences (e.g., $\delta_1=x_1-0, \delta_2=x_2-x_1, \ldots$). To recover the original identifiers during decompression, we need to compute a prefix sum
 ($x_i = \delta_i+x_{i-1}$). When using vectorized formats, we
 vectorize the computation of the prefix sum.
 All decoding times include differential coding to reconstruct the original integers.

We use a collection of posting lists extracted from the ClueWeb09 (Category B) data set. ClueWeb09 includes 50~million web pages. We have one posting list for each of the 1~million most frequent words---after excluding stop words and applying lemmatization. Documents are sorted lexicographically based on their URL prior to attributing document identifiers.\footnote{The posting-list data is freely available (\url{http://goo.gl/DygoQM}).}

We want to sort the posting lists by compressibility. For this purpose, the posting lists are grouped based on  length:
we store and process lists of lengths  $2^K$ to $2^{K+1}-1$ together for all values of $K$ needed by the corpus.
While the correlation is not perfect, shorter lists tend to be less compressible  than longer lists since their gaps tend to be larger.

We decode the compressed data sequentially to a  buffer of \num{4096} 32-bit integers (half the size of the \SI{32}{kB} L1~cache).
 For each group and each decoder, we compute the average decoding speed in billions of 32-bit integers per second (Bis). So that disk access is not an issue, the data is loaded in memory
prior to processing. The length of the arrays, their compressibility and the volume of data varies from group to group.

\begin{figure*}[htb]\centering
\subfloat[\label{fig:comp}Decoding speed in billions of integers per second (Bis) versus the compressibility of the data  (in bits per integer for VByte)]{
\includegraphics[width=0.47\textwidth]{{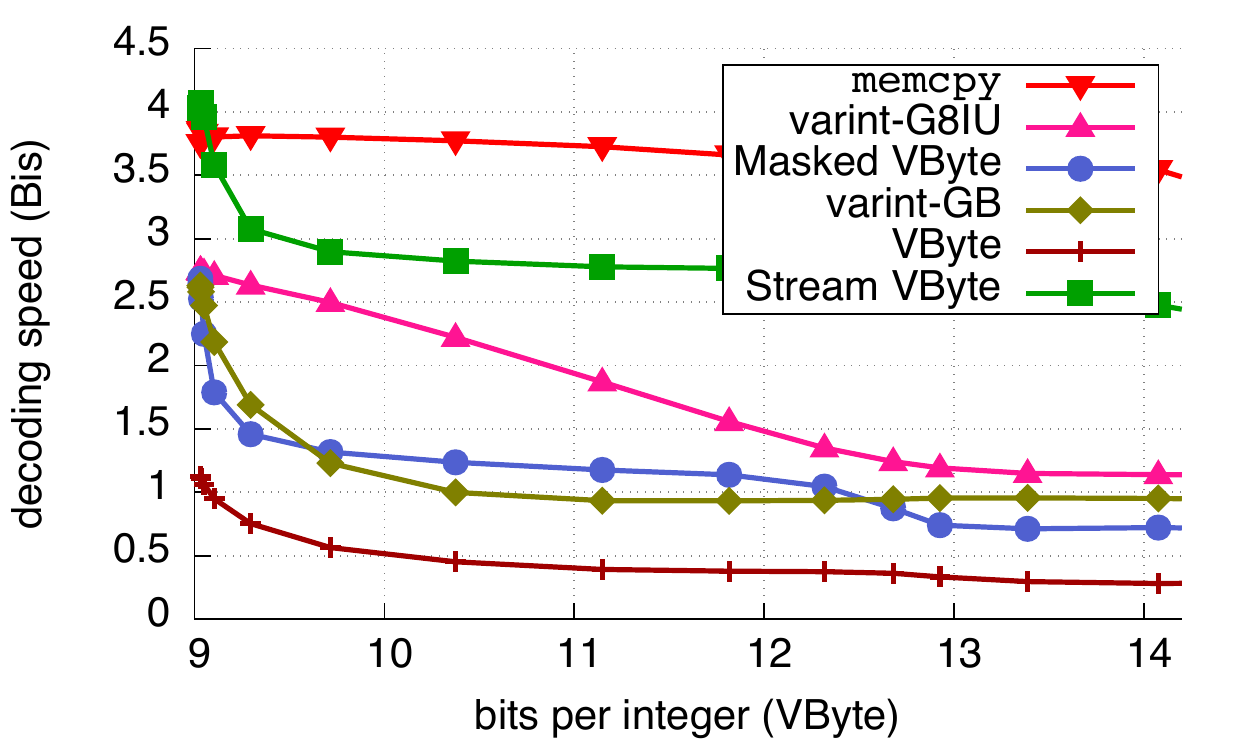}}

}\hfill{}
\subfloat[\label{fig:compsize}Extra storage space in bits per integer vs.\ the storage requirement of VByte]{
\includegraphics[width=0.47\textwidth]{{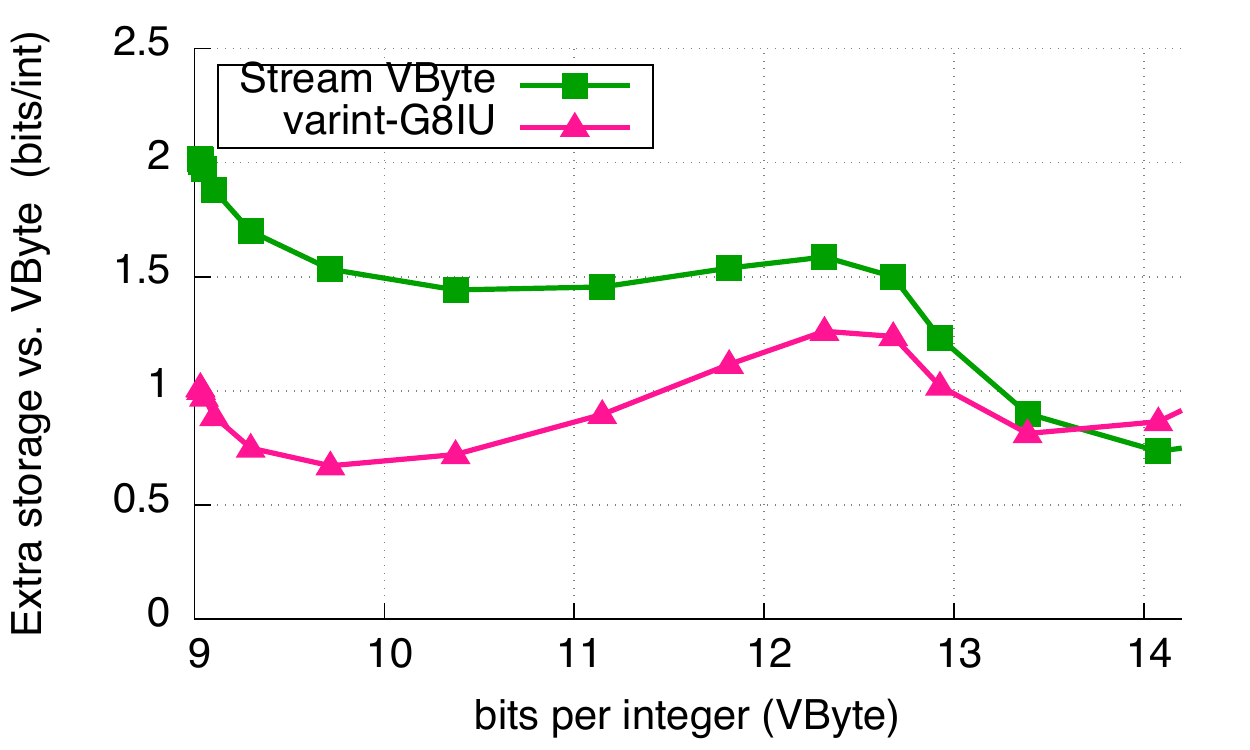}}
}
\caption{Results over sets of posting lists (ClueWeb)}
\end{figure*}

 Our results are summarized in Fig.~\ref{fig:comp}. 
 For \streamvbyte{}, the reported speed ranges from \SI{4.0}{Bis} for highly compressible data to \SI{1.1}{Bis} for less compressible data.
 The second fastest codec is \varintgiu{} with speeds ranging from
 \SI{2.7}{Bis} to \SI{1.1}{Bis}.
Next we have \maskedvbyte{} and  \varintgb{} with
speeds  ranging from
 \SI{2.6}{Bis} to \SI{0.5}{Bis}. Though \varintgb{}
 is more than \SI{50}{\percent} faster than \maskedvbyte{} when the data
 is poorly compressible, they are almost tied speed-wise when the data
 is more compressible.
 Finally, we find \vbyte{} with speed
 ranging from
 \SI{1.1}{Bis} to \SI{0.3}{Bis}.
For all groups of posting lists, \streamvbyte{} is fastest of the algorithms considered. It is always
at least $\approx 2.5\times$ faster than the conventional \vbyte{} decoder
 and sometimes nearly $5\times$ faster. Compared to \varintgiu{},
\streamvbyte{} can be twice as fast.
For reference, we also provide  the copy speed of the uncompressed
data by blocks of up to \num{4096}~integers using the C function \texttt{memcpy}.
For highly compressible data,  \streamvbyte{} is
  faster than \texttt{memcpy} because fewer bytes are read. In our worst case, \streamvbyte{} decompresses at \SI{70}{\percent} of the corresponding \texttt{memcpy} speed.

Though they are faster,  both \streamvbyte{} and  \varintgiu{}
use .5 to 2~more bits per integer than \vbyte{} in these tests: see Fig.~\ref{fig:compsize}.
\streamvbyte{}  and \varintgb{} have almost exactly the same
storage requirements due to their similar format.
Also \maskedvbyte{} and \vbyte{} have exactly
the same compressed format.

\begin{figure*}\centering
\subfloat[\label{fig:search}Seek speed]{
\includegraphics[width=0.47\textwidth]{{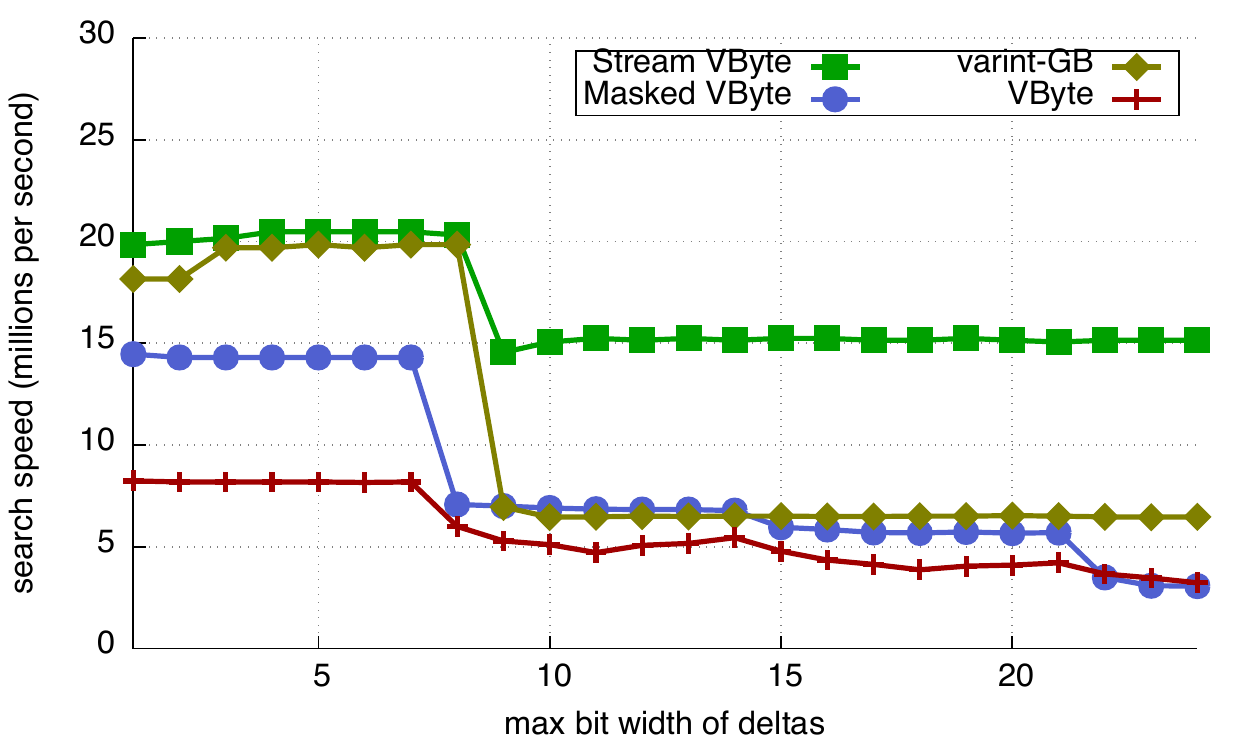}}

}\hfill{}
\subfloat[\label{fig:select}Select speed]{
\includegraphics[width=0.47\textwidth]{{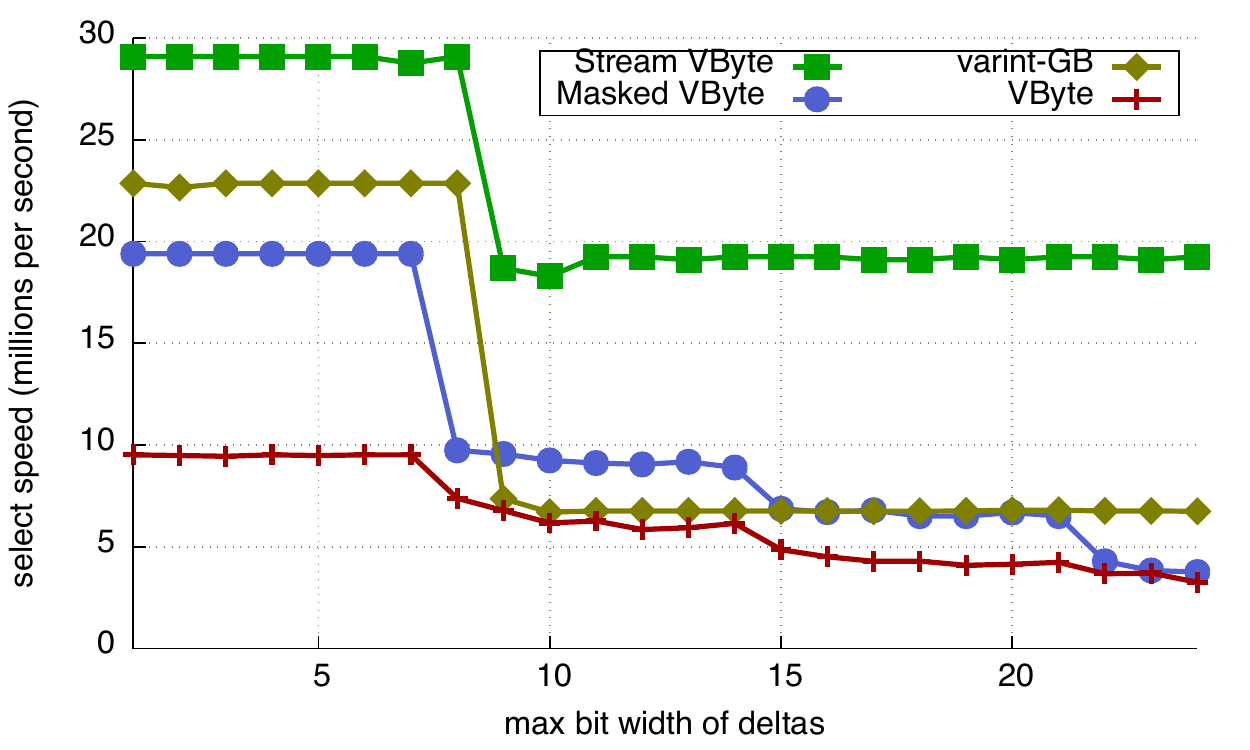}}
}
\caption{Seek and select results over blocks of 256 random integers}
\end{figure*}

Byte-oriented codecs make it convenient to program advanced operations directly
on the compressed stream. We benchmark two such operations: (1)~we seek the location of the first  value greater or equal to a given target, and retrieve this value, and (2)~we select  the $i^{\mathrm{th}}$~integer. 
Given a bit width $b \leq 24$, we first generate an array of 256~random integers in $[0,2^b)$: $\delta_1, \delta_2, \ldots$. The prefix sum is computed ($\delta_1, \delta_1+ \delta_2, \ldots $) and used as input data.  We omit \varintgiu{} in this experiment.
In Fig.~\ref{fig:search}, we randomly seek a value in range.
In Fig.~\ref{fig:select}, we randomly select the value at one of the indexes. In these tests,
 \streamvbyte{} offers the best speed and is up to three times faster than  \vbyte{}, with intermediate results for other codecs.
The performance is noticeably better when all deltas compress down to a single byte due the simple and predictable code path: it happens when all deltas fit in 7~bits for \vbyte{} and \maskedvbyte{}, and when they all fit in 8~bits for \varintgb{} and \streamvbyte{}.

\section*{Acknowledgments}

This work is supported by the National Research Council of Canada, under
grant 26143. We thank L. Boystov from CMU for sharing the posting list collection.

\bibliographystyle{model1-num-names}
\bibliography{varint}

\end{document}